\documentclass[]{cupconf}
\usepackage{graphicx}

\title[Metal-Rich A-Type Supergiants in M\,31]
      {Metal-Rich A-Type Supergiants in M\,31}
      
\author[N.~Przybilla et al.]{\\ Norbert Przybilla$^1$, Keith Butler$^2$ \& Rolf-Peter Kudritzki$^{3}$}
\affiliation{$^1$ Dr. Karl Remeis-Sternwarte Bamberg, Sternwartstrasse 7, D-96049 Bamberg, Germany \\ 
$^2$ Universit\"ats-Sternwarte M\"unchen, Scheinerstrasse 1, D-81679 M\"unchen, Germany\\ 
$^3$Institute for Astronomy, 2680 Woodlawn Drive, Honolulu, HI 96822, USA}

\begin{document}
\maketitle

\begin{abstract}
We discuss results of an exploratory non-LTE analysis of two metal-rich A-type
supergiants in M\,31. Using comprehensive model atoms we derive accurate 
atmospheric parameters from multiple indicators and show that non-LTE
effects on the abundance determination can be substantial (by a factor 2-3). 
The non-LTE analysis removes systematic trends apparent in the LTE approach
and reduces statistical uncertainties. Characteristic abundance patterns of the
light elements provide empirical constraints on the evolution of metal-rich
massive stars.
\end{abstract}

\firstsection
\section{Introduction}
Absorption by interstellar dust in the Galactic plane prohibits the
study of the Milky Way in its entirety. A comprehensive global view of a 
giant spiral galaxy can be obtained only for the nearest neighbour of this
class, the Andromeda galaxy (M\,31). The old stellar population of this, the
most luminous galaxy of the Local Group, turns out to be unexpectedly
metal-rich (see e.g. van den Bergh~1999) and present-day abundances as
traced by nebulae (H\,{\sc ii} regions, supernova remnants) also indicate
a metal-rich character and the presence of abundance gradients in the disk
(Dennefeld \& Kunth~1981;~Blair~et~al.~1982).

The current generation of large telescopes allows spectroscopy of luminous 
stars in~M\,31 to be carried out. Quantitative studies of massive blue supergiants 
(BSGs) can help to verify results from nebulae. In particular, analyses of
bright BA-type supergiants (BA-SGs) facilitate abundance determinations to be 
extended beyond the light and $\alpha$-process elements to iron group and s-process species.
This makes BA-SGs highly valuable for constraining the galactochemical
evolution of M\,31 empirically by tracing the abundance gradients. 
However, there is more to gain: 
observational constraints on the evolution of massive stars in a metal-rich
environment, using BSGs as probes for mixing with nuclear-processed matter, 
and the potential to employ
them as distance indicators via application of the flux-weighted
gravity-luminosity relationship (FGLR, Kudritzki et al.~2003).

Only a few studies of individual BSGs in M\,31 are available so far (Venn et al.~2000; Smartt et
al.~2001; Trundle et al.~2002). These are based either on the assumption of
LTE or on unblanketed non-LTE model atmospheres.
Here, we present results of an exploratory study of two A-SGs in M\,31.
A hybrid non-LTE analysis technique is used which considers line blanketing
and non-LTE line formation and is thus able to provide results of
hitherto unachieved accuracy and consistency (Przybilla~2002;~Przybilla~et~al.~2006).

\section{Observations and model computations}
Spectra of two luminous A-type supergiants
in the north-eastern arm of M\,31
were taken in November 2002 with {\sc Esi} 
on the Keck\,II telescope. The Echelle spectra were reduced with
the {\sc Makee} package and our own IDL-based routines for order merging and 
continuum normalisation. 
Complete wavelength coverage of the spectra between
$\sim$3900 and $\sim$9300\,{\AA} was obtained at high S/N ($>$\,150 in the
visual) and moderately high resolution ($R$\,$\simeq$\,8000). The dataset is
complemented by Keck\,I/{\sc Hires} spectra with a more limited wavelength
coverage ($R$\,$\simeq$\,35\,000, S/N$\sim$80) as utilised by Venn et al.~(2000). 

The model calculations are carried out in a hybrid non-LTE approach as
discussed in detail by Przybilla et al.~(2006). In brief, hydrostatic,
plane-parallel and line-blanketed LTE model atmospheres are computed with
{\sc Atlas9} (Kurucz~1993; with further modifications:
Przybilla et al.~2001). Then, non-LTE line formation is performed on the
resulting model stratifications. The coupled radiative transfer and
statistical equilibrium equations are solved and spectrum synthesis with refined
line-broadening theories is performed using {\sc Detail} and {\sc
Surface}. State-of-the-art non-LTE model atoms relying on data from {\em
ab-initio} computations, avoiding rough approximations wherever possible, 
are utilised for the stellar parameter and abundance determination.

\begin{table}
\caption[]{Atmospheric and fundamental stellar parameters with uncertainties}
\begin{center}
\begin{tabular}{lrrrrrrrrrr}
\noalign{\smallskip}
 Object & $T_{\rm eff}$\,(K) & $\log g$ & He & $[$M/H$]$ & 
$\xi$/$\zeta$/$v \sin i$\,(km/s) & $\log L$/$L_\odot$ & $M_{\rm e}$/$M_\odot$ &
$M_{\rm s}$/$M_\odot$ & $R$/$R_\odot$\\
\hline
\noalign{\smallskip}
41-3654 & 9200 & 1.00 & 0.13 & $+$0.13 & 8/20/36 & 5.63 & 29 & 24 & 257\\
(A2\,Iae) &  150 & 0.05 & 0.02 &    0.06 & 1/~\,5/~\,5 & 0.04 & 4 & 4 & 15\\
41-3712 & 8550 & 1.00 & 0.13 & $-$0.04 & 8/18/25 & 5.45 & 24 & 22 & 243\\
(A3\,Iae) &  150 & 0.05 & 0.02 &    0.05 & 1/~\,5/~\,5 & 0.04 & 3 & 4 & 14\\
\noalign{\smallskip}
\end{tabular}
\end{center}
\label{atm}
\end{table}

\section{Stellar parameters and elemental abundances}
The atmospheric parameters are derived spectroscopically from multiple
indicators, following the methodology described by Przybilla et al.~(2006).
Effective temperature~$T_{\rm eff}$ and surface gravity~$\log g$ 
are constrained from several non-LTE ionization equilibria 
(C\,{\sc i/ii}, N\,{\sc i/ii}, Mg\,{\sc i/ii}) and from modelling the
Stark-broadened profiles of the higher Bal\-mer and
Paschen lines. The internal accuracy of the method allows
the 1$\sigma$-uncertainties to be reduced to $\sim$1--2\% 
in~$T_{\rm eff}$ and to 0.05--0.10\,dex in~$\log g$. Several He\,{\sc i} lines are
used to derive the helium abundance. The stellar metallicity relative to the
solar standard $[$M/H$]$ (logarithmic values) is determined from the heavier metals (O, Mg, S,
Ti, Fe) in non-LTE. Microturbulent velocities
$\xi$ are obtained in the usual way by requiring abundances to be
independent of line equivalent width -- consistency is achieved from all
non-LTE species. Finally, macroturbulences $\zeta$ and rotational
velocities $v \sin i$ are determined from line profile fits. The results are
summarised in Table~\ref{atm}, where information on the fundamental
stellar parameters is also given: luminosity~$L$, evolutionary and spectroscopic 
mass~$M_{\rm e}$/$M_{\rm s}$ and radius~$R$. Photometric data of Massey et
al.~(2006) are adopted.

\begin{figure}
\includegraphics[width=0.995\linewidth]{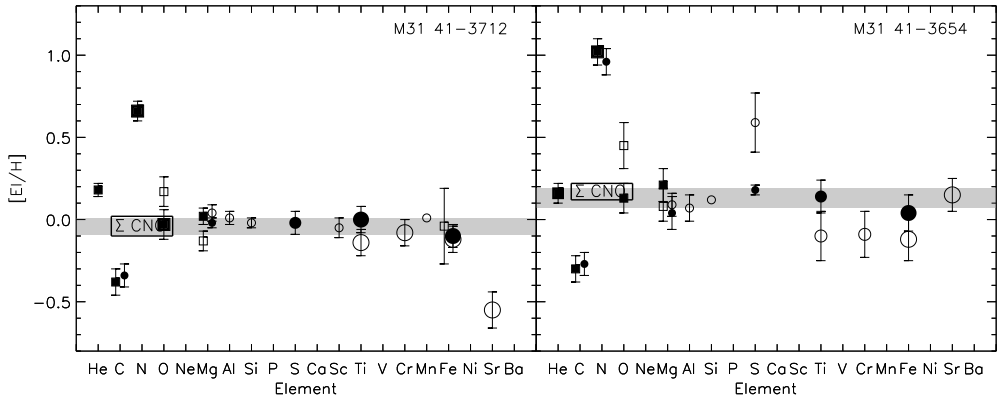}
\caption{Preliminary results from the elemental abundance analysis for our two sample
stars, relative to the solar composition (logarithmic scale, Grevesse \& Sauval~1998).
Filled symbols denote non-LTE, open
symbols LTE results. The symbol size codes the number of spectral lines
analysed -- small: 1 to 5, medium: 6 to 10, large: more than 10.
Boxes: neutral, circles: single-ionized species.
The error bars represent 1$\sigma$-uncertainties from the
line-to-line scatter. The grey shaded area marks the deduced metallicity of
the objects within 1$\sigma$-errors.
The non-LTE abundance analyses imply a scaled solar abundance distribution 
for the M\,31 objects. An exception are the light elements 
which have been affected by mixing with nuclear-processed matter.}
\label{fig1}
\end{figure}

Elemental abundances are determined for several chemical species, 
with many of the astrophysically most interesting in non-LTE and the
remainder in LTE, see Fig.~\ref{fig1}. 
The two M\,31 supergiants are more metal-rich than the
Galactic BA-SGs studied using the same method (Przybilla et al.~2006; Firnstein \&
Przybilla~2006; Schiller \& Przybilla~2006), by up to $\sim$0.2\,dex. 
One object is found to show super-solar metallicity.
The abundance distribution of the heavier elements follows a scaled solar
pattern while the light elements have been affected by mixing
with CN-cycled matter.

\section{Results and discussion}
The non-LTE computations
reduce random errors and remove systematic trends in the analysis.
Inappropriate LTE analyses tend to systematically underestimate iron group
abundances and overestimate the light and $\alpha$-process element abundances 
by up to factors of 2-3 (most notable for M31-41-3654, while M31-41-3712 is
less affected). This is because of the different
responses of these species to radiative and collisional processes in the 
microscopic picture, which is explained by fundamental differences of their 
detailed atomic structure. This is not taken into account in LTE. 
Contrary to common assumptions, significant non-LTE abundance corrections of
$\sim$0.3\,dex can be found even for the weakest lines
($W_{\lambda}$\,$\sim$\,10\,m{\AA}). Non-LTE abundance uncertainties
amount to typically 0.05--0.10\,dex (random) and $\sim$0.10\,dex
(systematic 1$\sigma$-errors). Note that line-blocking
effects increase with metallicity, such that photon
mean-free paths are reduced in metal-rich environments and the non-LTE effects
correspondingly. This is the reason why non-LTE effects in
these objects close to the Eddington limit are similar to those in
less-extreme Galactic BA-type supergiants with -- on average -- lower metallicity.

\begin{figure}
\centering\includegraphics[width=0.67\linewidth]{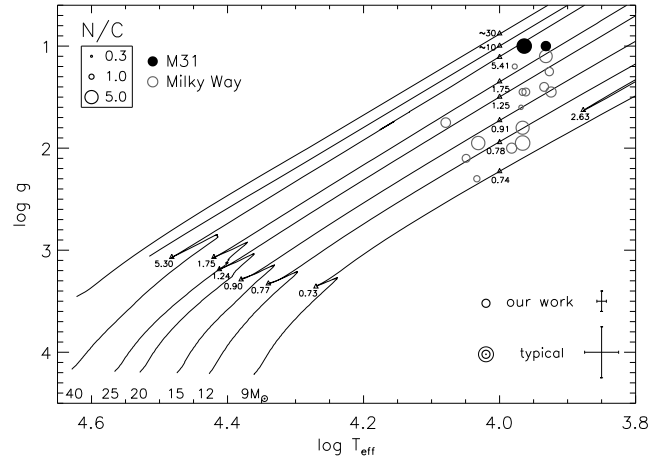}\\[-4mm]
\caption{Observational constraints on massive star evolution: N/C ratios as
tracers of mixing with nuclear-processed material. Displayed are evolution
tracks for rotating stars ($v_{\rm ini}\,=\,300$\,km/s, with marks
indicating N/C ratios from the models: Meynet \& Maeder~2003)
at solar metallicity. The model computations adopt an initial N/C\,$\sim$\,0.3. 
Observed N/C ratios (typical values are
indicated in the box) for a sample of Galactic BA-SGs
(Przybilla et al.~2006; Firnstein \& Przybilla~2006; Schiller
\& Przybilla~2006) and for
the two (slightly) metal-rich M\,31 objects of the present work are
indicated, all analysed in a homogeneous way. Error bars as characteristic for our 
work and for similar studies from the literature are~given.}
\label{fig2}
\end{figure}

Fundamental stellar parameters and light element abundances 
allow us to discuss the two M\,31 objects in the context of stellar evolution. The
comparison with evolutionary tracks for rotating massive stars is made in
Fig.~\ref{fig2}, which also summarises results from a Galactic sample of
BA-SGs. The M\,31 supergiants extend the sample towards higher luminosities/stellar
masses than possible in the Galactic study, 
and towards higher metallicity. Both objects appear to cross
the Hertzsprung-Russell diagram towards the red supergiant stage for the
first time, because of the absence of extremely high helium abundances and N/C ratios expected
for stars entering the Wolf-Rayet phase. The predicted
trend of increased chemical mixing (strong N and moderate He enrichment, C
depletion and almost constant O as a result of the action of the CNO-cycle and
transport to the stellar surface because of meridional circulation and
dynamical instabilities) with 
increasing stellar mass is qualitatively recovered. Note a group of highly 
processed stars at $M_0$\,$<$\,15\,M$_\odot$, which suggests an
extension of blue loops towards higher temperatures than predicted. The
observed N/C ratios are generally higher than indicated by theory. Stellar
evolution computations accounting for the interplay of rotation and magnetic
fields (e.g. Maeder \& Meynet~2005) may resolve this discrepancy as they
predict a much higher efficiency for chemical mixing. Also the recent
revision of the cross-section for the bottleneck reaction
$^{14}$N(p,$\gamma$)$^{15}$O in the CN-branch of the CNO-cycle  
by almost a factor~2 (Lemut et al.~2006) will be of importance. 

Finally, improved stellar parameter allow for a
re-evaluation of the two M\,31 supergiants in the empirical calibration of
the FGLR (Kudritzki et al.~2003). Two factors play a r\^ole in this context:
a revision of the previously used photometric data of Magnier et al.~(1992),
which has been shown to suffer from systematic uncertainties (Massey et
al.~2006), and the extended wavelength coverage (and high S/N) of the {\sc
Esi} spectra, which allows the atmospheric parameters to be constrained more
precisely. As a consequence some of the largest deviations from the
empirical relation can be explained.



\begin{thebibliography}{99}

\bibitem[]{Blairetal1982}
        Blair, W.~P., Kirshner, R.~P. \& Chevalier, R.~A. (1982).
	\textit{ApJ} \textbf{254}, 50-69.

\bibitem[]{DeKu1981}
        Dennefeld, M. \& Kunth, D. (1981).
	\textit{AJ} \textbf{86}, 989-997.

\bibitem[]{FiPr06}
        Firnstein, M. \& Przybilla, N. (2006).
	\textit{Proceedings of Science}, PoS(NIC-IX)095.

\bibitem[]{GrSa98}
        Grevesse, N. \& Sauval, A.J. (1998).
	\textit{Space Sci. Rev.} \textbf{85}, 161-174

\bibitem[]{Kudritzkietal2003}
        Kudritzki, R.~P., Bresolin, F. \& Przybilla, N. (2003).
	\textit{ApJ} \textbf{582}, L83-L86

\bibitem[]{Kurucz1993}
        Kurucz, R.L. (1993). 
	Kurucz CD-ROM No. 13 (Cambridge, Mass.: SAO).

\bibitem[]{Lemutetal2006}
        Lemut, A., Bemmerer, D., Confortola, F. et al. (2006).
	\textit{Physics Letters B} \textbf{634}, 483-487.

\bibitem[]{MaMe2005}
        Maeder, A. \& Meynet, G. (2005).
	        \textit{A\&A} \textbf{440}, 1041-1049.

\bibitem[]{MeMa2003}
        Meynet, G. \& Maeder, A. (2003).
	\textit{A\&A} \textbf{404}, 975-990.

\bibitem[]{Magnieretal1992}
        Magnier, E.~A., Lewin, W.~H.~G., van Paradijs, J. et al. (1992).
	\textit{A\&AS} \textbf{96}, 379-388.

\bibitem[]{Masseyetal2006}
        Massey, P., Olsen, K.~A.~G., Hodge, P.~W., Strong, S.~B. et al. (2006). 
	\textit{AJ} \textbf{131}, 2478-2496.

\bibitem[]{Przybilla2002}
	Przybilla, N. (2002).
	Ph.\,D. Thesis, University Munich.
	
\bibitem[]{Przybillaetal2001}
        Przybilla, N., Butler, K. \& Kudritzki, R.P. (2001).
	\textit{A\&A} \textbf{379}, 936-954.
	
\bibitem[]{Przybillaetal2006}
	Przybilla, N., Butler, K., Becker, S.R. \& Kudritzki, R.~P.~(2006).
	\textit{A\&A} \textbf{445}, 1099-1126.

\bibitem[]{SchPr06}
        Schiller, F. \& Przybilla, N. (2006).
	\textit{Proceedings of Science}, PoS(NIC-IX)174.

\bibitem[]{Smarttetal2001}
        Smartt, S.~J., Crowther, P.~A., Dufton, P.~L. et al. (2001).
	\textit{MNRAS} \textbf{325}, 257-272.

\bibitem[]{Trundleetal2002}
        Trundle, C., Dufton, P.~L., Lennon, D.~J. et al. (2002). 
	\textit{A\&A} \textbf{395}, 519-533.

\bibitem[]{vdB1999}
        van den Bergh, S. (1999).
	\textit{A\&AR} \textbf{9}, 273-318.

\bibitem[]{Vennetal2000}
        Venn, K.~A., McCarthy, J.~K., Lennon, D.~J. et al. (2000). 
	\textit{ApJ} \textbf{541}, 610-623.

\end{thebibliography}
\end{document}